\documentclass[reprint, amsmath,amssymb, aps]{revtex4-2}

\usepackage{graphicx}
\usepackage{dcolumn}
\usepackage{bm}
\usepackage{xcolor}
\usepackage{float}
\usepackage{amsmath}
\usepackage{mathtools}
\usepackage{tabularx}
\usepackage{diagbox}
\usepackage{hhline}

\begin{document}

\title{Fast but multi-partisan: Bursts of communication increase opinion diversity in the temporal Deffuant model}

\author{F. Zarei}
\author{Y. Gandica}
\author{L. E. C. Rocha}
\email{luis.rocha@ugent.be}
\affiliation{Department of Economics, Ghent University, Ghent, Belgium}
\affiliation{Department of Physics and Astronomy, Ghent University, Ghent, Belgium}
\affiliation{Department of Mathematics, Valencian International University, Valencia, Spain}

\date{\today}

\begin{abstract}
Human interactions create social networks forming the backbone of societies. Individuals adjust their opinions by exchanging information through social interactions. Two recurrent questions are whether social structures promote opinion polarisation or consensus in societies and whether polarisation can be avoided, particularly on social media. In this paper, we hypothesise that not only network structure but also the timings of social interactions regulate the emergence of opinion clusters. We devise a temporal version of the Deffuant opinion model where pairwise interactions follow temporal patterns and show that burstiness alone is sufficient to refrain from consensus and polarisation by promoting the reinforcement of local opinions. Individuals self-organise into a multi-partisan society due to network clustering, but the diversity of opinion clusters further increases with burstiness, particularly when individuals have low tolerance and prefer to adjust to similar peers. The emergent opinion landscape is well-balanced regarding clusters' size, with a small fraction of individuals converging to extreme opinions. We thus argue that polarisation is more likely to emerge in social media than offline social networks because of the relatively low social clustering observed online. Counter-intuitively, strengthening online social networks by increasing social redundancy may be a venue to reduce polarisation and promote opinion diversity.
\end{abstract}


\maketitle

\section{Introduction}

Individual opinions are created by combining self-reflection and external inputs, such as mass media and information from social interactions. Social interactions play a significant role in opinion dynamics because individuals might be influenced by their peers (i.e.\ social contacts), and consequently, opinions diffuse through social networks~\cite{brown2007, brown2007word, muchnik2013social, state2015, delvicario2016}. In the last decades, social media has become widespread by providing platforms for online social networks. In contrast to offline social networks, those networks are not bound by the physical space and thus increase people's exposure to information via long-range connections. The broader exposure facilitates the reinforcement of certain opinions with the formation of echo chambers, which eventually leads to polarisation and radicalisation of ideas~\cite{del2016echo, delvicario2016, mason2018uncivil, jaaskelainen2020online, conover2011political, mocanu2015collective, gargiulo2017role, quattrociocchi2016echo}. Such phenomena may be exacerbated by sharing emotional news stories, controversial opinions~\cite{mocanu2015collective, conover2011political, quattrociocchi2016echo}, or by controlling how information reaches social contacts~\cite{perra2019modelling, Woolley2017}.

There are different paradigms to model opinion dynamics~\cite{castellano2009, Xia2011, guilbeault2018complex, cencetti2023}. An important class of models is based on reinforcing opinions by social contacts~\cite{sznajd2000, galam2005, centola2010}. Such models assume that individuals are more likely to change their opinions if enough neighbours have a particular opinion. This complex contagion mechanism has been observed in experimental settings and social media and shown to depend on the underlying social network structure~\cite{lerman2010information, centola2010, state2015}. The structural heterogeneity of complex social networks plays a pivotal role in regulating the spread of opinions via complex contagion. For example, clustering (e.g.\ network triangles) creates redundancy and locally reinforces the prevalent opinion within the group (echo chambers)~\cite{centola2010, perra2019modelling}. On the other hand, socially poorly connected individuals will constantly be exposed to a single opinion and thus less to reinforcement. Another class of opinion dynamic models do not implicitly assume that reinforcement is needed, but single exposure may be sufficient~\cite{deffuant2000mixing, Hegselmann2002, Redner2019}. Some of these models make the realistic assumption that, despite individuals being embedded in social networks, interactions are most often one-to-one and thus, every social interaction has the potential to contribute to adjusting someone's opinion. The Deffuant model captures these dynamics, with individuals adjusting their opinions (defined by real values from 0 to 1) at each pairwise social interaction if the difference between their opinions is less than a given threshold~\cite{deffuant2000mixing}. One consequence of such a mechanism is that reinforcement is weakened because, at each interaction, a single individual influences only another individual, and thus some network structures become less relevant~\cite{Fortunato2005}.

Another aspect of opinion dynamics is the timings of interaction events between individuals. Temporal patterns of human communication follow some regularity, for example, circadian or weekly cycles, but are mostly highly heterogeneous. A common temporal heterogeneity is burstiness, i.e.\ bursts of interaction events followed by periods of inactivity~\cite{barabasi2005origin, goh2008burstiness, rocha2011, karsai2011}. burstiness is characterised by a right-skewed distribution of inter-event times between subsequent events and has been observed in various forms of social media and human communication~\cite{barabasi2005origin}. Such temporal activity affects dynamic processes, e.g.\ epidemic spread and opinion dynamics~\cite{rocha2013bursts, karsai2011, jo2021, li2023}, and together with network structure, regulate the relaxation (mixing) time to stationarity~\cite{delvenne2015diffusion}.

In this paper, we hypothesise that peer pressure (complex contagion) is unnecessary to reinforce opinions and that reinforcement can be obtained by repeated exposure to the same social contact. We thus design a variation of the Deffuant model (originally based on a single pairwise interaction per time step and therefore unable to create peer pressure) where the interactions follow burst activity patterns. Our model shows that burstiness not only slows down the dynamics towards stationarity but also increases the number of opinion clusters, i.e.\ promotes a multi-partisan society, independently of the network structure. Diversity increases the likelihood of extreme opinions but does not lead to the emergence of disproportionate large clusters. Furthermore, structurally clustered networks boost those effects by combining two forms of reinforcement, the first due to repeated exposure to the same social contact (temporal) and the second by the collective impact of peer pressure (structural).

\section{Materials and Methods}
\label{sec:Model}

\subsection{Social Networks}

A social network comprises $N$ nodes, each node $i$ representing an individual $i$, and $E$ links $(i,j)$ representing social ties between nodes (individuals) $i$ and $j$. The degree $k_i$ is the number of the social relations to node $i$, whereas the average number of social ties in the network is given by the average degree $\langle \kappa \rangle$. Nodes are clustered if they form triangles (i.e.\ node $A$ is connected to $B$ and $C$, which are in turn connected as well), and the level of clustering can be measured by the clustering coefficient $cc_i = 2 e_i / (k_i (k_i-1))$. At the mesoscale, clustering can be quantified by the modularity $Q$, which measures the level of connectivity within groups of nodes (i.e.\ network communities) compared to what would be expected by chance. Higher modularity indicates a more robust community structure. These types of clustering will be referred to as structural clustering. Degree assortativity $r$ is the tendency of nodes with similar degrees to be connected (homophily by degree)~\cite{newman2018networks}.

Three network models are used to create social ties (links) between individuals. The Erd\"os-R\'enyi (ER) is the reference network model in which social ties are formed between pairs of nodes with a fixed probability $p$. This model generates a homogeneous network where nodes have a characteristic degree~\cite{newman2018networks}. The second model is the Watts-Strogatz (WS), which generates networks with local clustering (high clustering coefficient) yet short connections between pairs of nodes. The WS model is built by connecting the $k_{nn}=6$ nearest neighbours of a node and then rewiring the social ties with probability $q$~\cite{newman2018networks}. The third model (fitness model) reproduces social networks more realistically by assuming that individuals tend to form social ties with others similar to them according to an attribute, e.g., age or gender (homophily by attribute). In the model, nodes are added to a network following a preferential attachment mechanism regulated by the level of similarity between the individuals. This is incorporated by using a fitness function $\phi_\text{N}(i)$ such that a newly added node $N$ preferentially connects to a high-degree node taking into account its level of similarity to the existing nodes (eq.~\ref{eq:phi}).

\begin{equation}
  \label{eq:phi}
  \phi_\text{N(i)} \propto k_\text{i} \exp\left( -\beta \left| \theta_\text{N}-\theta_\text{i} \right|\right),
\end{equation}
where $k_\text{i}$ and $\theta_\text{i}$ are respectively the degree and attribute of an existing node $i$, $\theta_\text{N}$ is the attribute of the newly added node $N$, and $\beta$ is a coefficient controlling the fitness level regarding the attribute.

When $\beta=0$, the growth mechanism follows the classical preferential attachment, resulting in a Barabasi-Albert network with degree distribution $P(k)\propto k^{-3}$~\cite{newman2018networks}. Attribute similarity is introduced with $\beta > 0$, which competes with the degree to attract new links. In the limit of $\beta \to \infty$, $\phi \to 0$, and the distribution becomes closer to an exponential. This fitness function thus enables sweeping connections from high-degree nodes (reducing the rich-get-richer effect) to those with similar attribute values (balancing the network towards attribute homophily).
    
\subsection{Dynamic Social Networks}
\label{dy_soc_net}

A dynamic social network is defined by a sequence of links $(i,j)$ active at certain times $t$. In other words, a link $(i,j)$ can be active or inactive at time $t$, unlike static networks where links are persistently active, i.e.\ independently of the time. To generate dynamic networks following specific temporal patterns, we first define a fixed network structure and then assign the next activation time of each link using $t_{next} = t+\Delta t$, where $\Delta t$ is the inter-event time sampled from a distribution ($P(\Delta t)$) and $t$ is the time of the current link activation~\cite{rocha2013bursts, masuda2018gillespie}. This procedure guarantees uncorrelation of structure and timings of link activation, where the dynamic network contains a pre-defined topology with asynchronous activation of links to the same node, also following a pre-defined activation pattern.

In this computational model, links are initially active at random times $t$. The network evolves by sampling a new $\Delta t$ from the inter-event time distribution each time a link $(i,j)$ becomes active. The subsequent activation time of the same link $(i,j)$ is thus $t_{next} = t + \Delta t$. Links are sorted chronologically to guarantee that the continuous activation times are ordered. The Markov process results in a transient period discarded before the network dynamics becomes stationary~\cite{rocha2013bursts, masuda2018gillespie}.

Different inter-event time distributions $P(\Delta t)$ can be used to simulate chosen patterns. The baseline model is the exponential distribution, $P(\Delta t) = b e^{-b \Delta t }$, with $\langle \Delta t \rangle = 1/b$, corresponding to a memoryless Poisson process. This model generates homogenous activation times and is equivalent to activating a link uniformly at random at each time step $t$. Empirical evidence, however, suggests that human interactions have memory and are better described by heterogeneous right-skewed distributions. We use a log-normal distribution of inter-event times to capture the burstiness of link activations (eq.~\ref{eq:iet_dist}).

\begin{equation}
    \label{eq:iet_dist}
    P(\Delta t) = \frac{1}{\sqrt{2\pi} \sigma \Delta t} e^{-\frac{(\ln \Delta t - \nu)^2}{2 \sigma ^2} },
\end{equation}

where $\langle \Delta t \rangle = \exp(\nu+\sigma^2/2)$ ($\nu$ is a parameter) and the burstiness depends on the variance $\sigma$ of the distribution~\cite{goh2008burstiness, delvenne2015diffusion}. The log-normal distribution approaches a Dirac delta distribution as $\sigma \to 0$ and has nearly linear log density as $\sigma \gg 1$ (a power-law has linear log density). The log-normal distribution provides a convenient transition function between random or low burstiness ($\sigma \ll 1$) and nearly power-law or high burstiness ($\sigma \gg 1$)~\cite{delvenne2015diffusion}.

\subsection{Temporal Deffuant Model}

The Deffuant model of opinion dynamics assumes a population of $N$ individuals, wherein individuals $i$ and $j$ are connected via a static social tie $(i,j)$~\cite{deffuant2000mixing}. The collection of such social ties forms a social network. The opinions of the two individuals at time $t$ are represented respectively by $x_\text{i}(t)$ and $x_\text{j}(t)$, with $x(t) \in [0,1]$. Each individual $i$ is assigned a random opinion $x_\text{i}(t)$ at time $t=0$. At each time step $t$, a randomly chosen pair of connected individuals $(i,j)$ interact pairwise. The interaction is successful if the difference between the individuals' opinions is smaller than a confidence level $d$ (eq.~\ref{eq1}), where $d$ is constant and corresponds to the individuals' openness to adapt their opinion.

\begin{eqnarray}
    |x_\text{i}(t)-x_\text{j}(t)|<d
    \label{eq1}
\end{eqnarray}

A successful interaction leads to both individuals updating their opinions based on the difference between their original opinions (eq.~\ref{eq2}). The parameter $\mu \in [0,1]$ defines the influenceability of an individual towards another, i.e.\ the extent that two opinions can adjust and converge.

\begin{eqnarray}
    x_\text{i}(t+1)&&= x_\text{i}(t)+ \mu (x_\text{j}(t) - x_\text{i}(t) ) \nonumber\\
    x_\text{j}(t+1)&&= x_\text{j}(t)+ \mu (x_\text{i}(t) - x_\text{j}(t) )
    \label{eq2}
\end{eqnarray}

We design a variation of this model incorporating temporal dynamics to the original Deffuant model via dynamic social networks, meaning that a social tie can now be active or inactive at different times. The opinions of both individuals $x_\text{i}(t)$ and $x_\text{j}(t)$ can thus only be updated if the tie $(i,j)$ is active at time $t$. The activation times follow independent (asynchronous) temporal activity patterns. The dynamic network is initialised for a chosen inter-event time distribution, and after the initial transient, the Deffuant opinion dynamics start, and updates follow the sequence of link activation (See~\ref{dy_soc_net}). The opinion dynamics evolve until the stabilisation of the opinions. This stabilisation time $T_\text{f}$ is defined as the time after $\Delta t=N$ consecutive time steps without a successful exchange of opinions, i.e.\ either the nodes converged to the same opinion, or the opinions became so different that no more updates are possible. In this stationary state, individuals are clustered according to their opinions (hereafter called opinion clusters). This is done by sorting the (values of) opinions of the individuals. If two successive opinions differ less than $\epsilon=10^{-4}$, the individuals are assigned to the same opinion cluster. Otherwise, a new opinion cluster is created. The number and sizes of opinion clusters are given by $N_\text{f}$ and $S_\text{f}$, respectively. Some individuals may get stuck on a particular opinion and not converge to an opinion cluster. To distinguish those cases and individuals collectively forming opinion clusters, an opinion cluster must have at least $1\%$ of the total population $N$.

The convergence parameter $\mu$ in the Deffuant model influences the relative speed of the dynamics but not the final state (i.e.\ the number of opinion clusters)~\cite{deffuant2000mixing}. Therefore, we set $\mu=0.5$ in our experiments. We also set $N=1000$ and $\langle \kappa \rangle=20$. For each experiment, we generate five independent samples of a chosen random network model with the same set of parameters and repeat the simulations ten times with random start conditions. Therefore, the average and standard deviations are calculated over $m=50$ points.

\begin{figure*}[!th]
    \centering
    \includegraphics[scale=0.5]{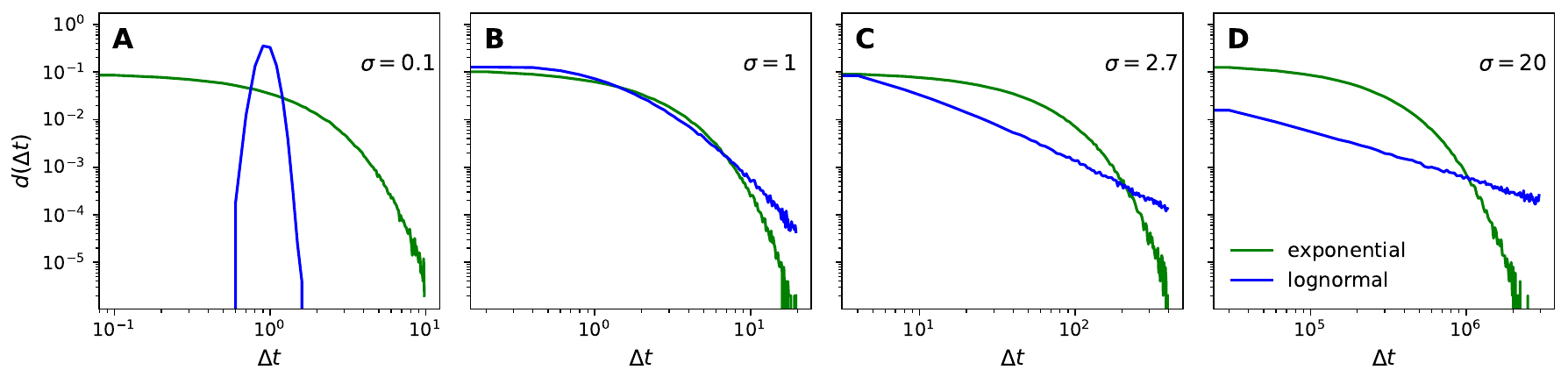}
    \centering\caption{Exponential and log-normal inter-event time distributions for (A) $\sigma=0.1$ ($\langle \Delta t \rangle = 1.0$), (B) $\sigma=1$ ($\langle \Delta t \rangle = 1.65$), (C) $\sigma=2.7$ ($\langle \Delta t \rangle = 37.9$), and (D) $\sigma=20$ ($\langle \Delta t \rangle = 1.9 \times 10^4$). The parameters of the distributions are set such that both distributions have the same mean inter-event times. Both axes are in log scale.}
    \label{fig01}
\end{figure*}

\section{Results}

\subsection{Evolution of the Opinion Dynamics}

Figure~\ref{fig01} shows the log-normal distribution with different levels of burstiness, together with a reference exponential distribution with the same average inter-event time. A low value of $\sigma$ gives a distribution peaked around the mean (which results in nearly regular activation times). In contrast, increasingly right-skewed distributions (and thus increasingly burst activity) are obtained for increasing $\sigma$. 

\begin{figure*}[!th]
   \includegraphics[scale=0.4]{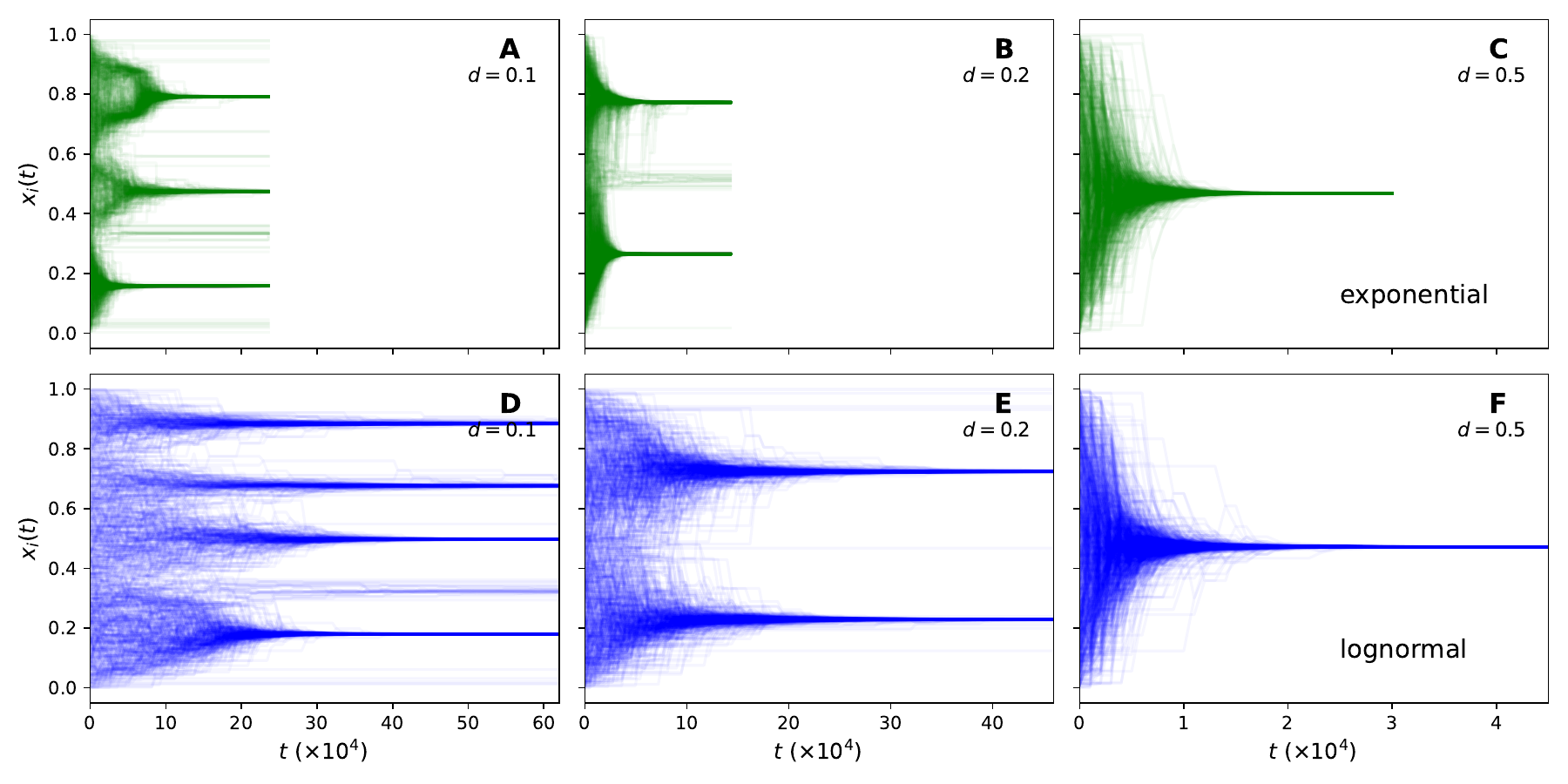}
    \centering\caption{Temporal evolution of individual opinions for varying values of $d$, using (A-C) exponential and (D-F) log-normal ($\sigma=2.7$) inter-event time distributions, with same $\langle \Delta t \rangle = 37.9 \rangle$. The underlying network structure follows the Erd\"os-R\'enyi network model.}
    \label{fig02}
\end{figure*} 

The evolution of the opinion dynamics with exponential inter-event times on a random network is equivalent to the standard Deffuant model on static networks~\cite{deffuant2000mixing}. Initially, opinions are homogeneously distributed among the individuals, but self-organisation leads to the emergence of opinion clusters in the stationary state. The number of opinion clusters and the stabilisation time depends on the confidence level $d$, with lower tolerance (i.e.\ lower values of $d$), leading to more opinion clusters (less consensus) (Fig.~\ref{fig02}A-C). Burst activation patterns (log-normal distribution) also decrease consensus as $d$ decreases. However, the system generally takes longer to reach the stationary state, and the number of emergent opinion clusters changes, for some configurations, compared to the reference exponential case using the same average inter-event times (Fig.~\ref{fig02}D-F). Some individuals do not change opinions through social interactions due to poor connectivity. Consequently, their opinions do not converge to an opinion cluster. This effect decreases with burst activity.

\begin{figure*}[!th]
    \centering
    \includegraphics[scale=0.35]{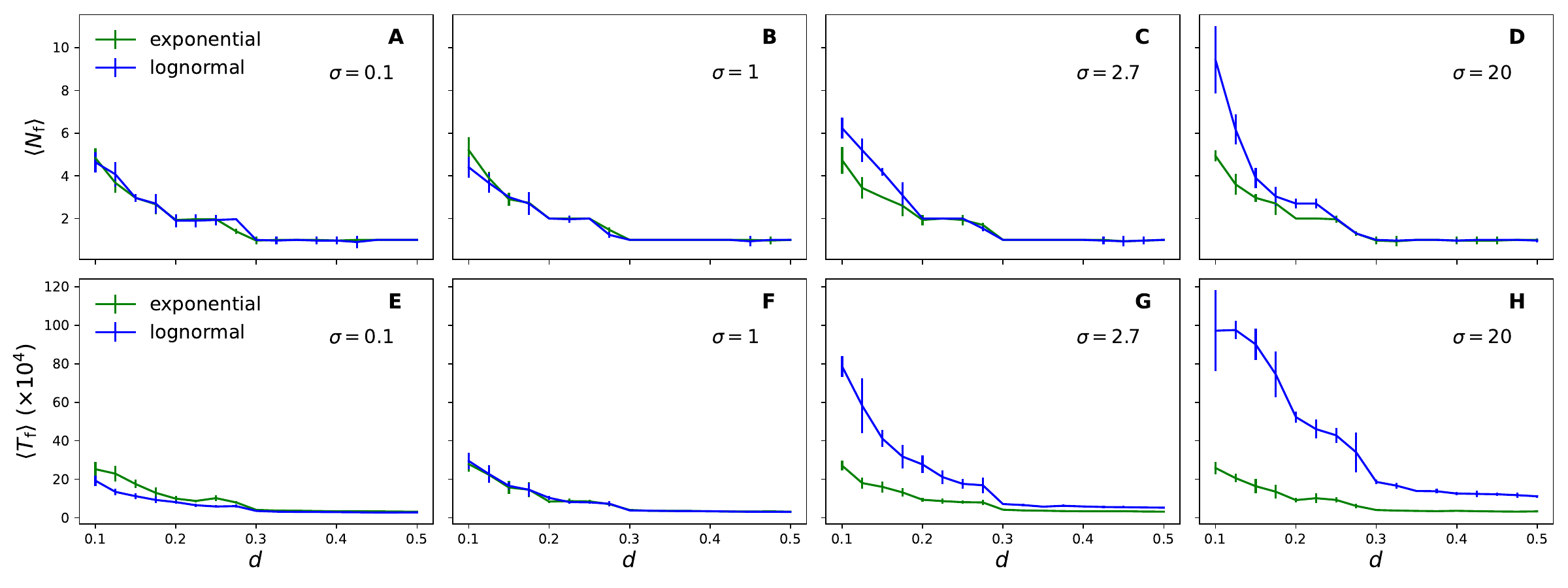}
    \centering\caption{(A-D) The average number of opinion clusters ($\langle N_\text{f} \rangle$) and (E-H) the average stabilisation time ($\langle T_\text{f} \rangle$) at the stationary state for different levels of burstiness ($\sigma$) as a function of the confidence level $d$. The averages are taken over $m=50$ realisations. Vertical bars represent the standard errors. }
    \label{fig03}
\end{figure*}

Figure~\ref{fig03}(A-D) shows the average number of final opinion clusters ($\langle N_{\text{f}} \rangle$) for various levels of burstiness (log-normal with $\nu=0$ and a given $\sigma$). A single opinion cluster always emerges for confidence levels $d \geq 0.3$, like the homogeneous (exponential) and non-temporal cases (standard Deffuant model). This indicates that a sufficiently high tolerance for different opinions leads individuals to self-organise towards consensus. On the other hand, for $d < 0.3$, the population is split into multiple opinion clusters. Near regular ($\sigma=0.1$) and exponential temporal patterns show similar results, but increasing heterogeneity (i.e.\ burstiness, with larger $\sigma$) leads to an increasing number of opinion clusters (more diversity). For example, $\sigma=20$ creates two times more opinion clusters ($\langle N_\text{f} \rangle \sim 9$, for $d=0.1$) than what would be created in the homogeneous case with the same $\langle \delta t \rangle$ and $d$. Furthermore, a significant difference in the number of opinion clusters happens only for $d=0.1$ (low tolerance) if $\sigma=1$. A relatively small increase in burstiness ($\sigma=2.7$) increases the tolerance level ($d=0.2$), in which bursts disproportionally affect the number of emerging opinion clusters.

Figure~\ref{fig03}(E-H) shows that the stabilisation times are also affected by burstiness; for both temporal patterns, decreasing $d$ increases $\langle T_\text{f} \rangle$. Lower $d$ means lower tolerance, which results in fewer interactions leading to opinion updates and thus requiring longer times before reaching stationarity. For $d<0.3$, near regular ($\sigma=0.1$) interactions lead to a small but significant speedup in the convergence to stationarity compared to the homogeneous case. On the other hand, increasing burstiness (larger $\sigma$) leads to a substantial slowdown (up to 5 times slower for $\sigma=20$). In contrast to the number of opinion clusters, burstiness always affects the convergence times (compared to the homogeneous case); higher burstiness leads to a dynamic slowdown, despite both temporal patterns leading to global consensus for confidence levels $d>0.3$.

burstiness of social interactions means that pairs of individuals interact often during specific short periods of time and those interactions remain inactive for extended times. Such fast and repeated interactions reinforce existing opinions, increasing the similarity of opinions locally and consequently creating spots or local clustering of opinions (local consensus). This local consensus makes the opinion clusters diverge more quickly early in the dynamics, followed by a slowdown in the convergence to stationarity due to the long periods of social tie inactivity (i.e.\ the large inter-event times).

\subsection{Structure, burstiness and Diversity}

\begin{figure*}[!th]
    \centering
      \includegraphics[scale=0.55]{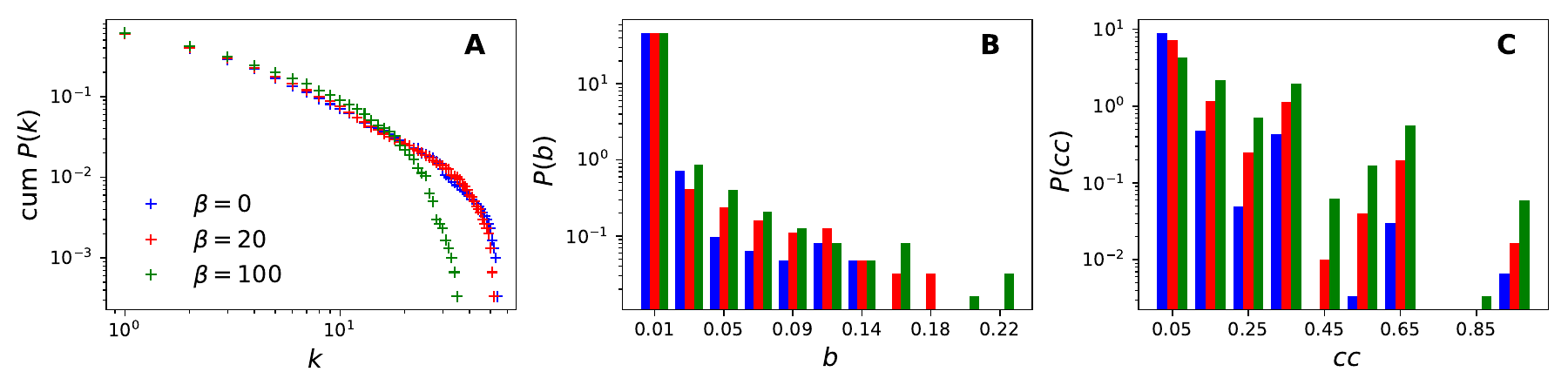}
    \centering\caption{Structural characteristics of the fitness random network model. (A) cumulative degree distribution, (B) betweenness centrality distribution (bin size 0.02), and (C) clustering coefficient distribution (bin size 0.1) for different values of the fitness parameter $\beta$. The histograms (B) and (C) are normalised probability density functions, where each bin's height is scaled to maintain the total area under the histogram as 1.} 
    \label{fig04}
\end{figure*}

The complex network structure is included in the analysis via the fitness random model (See Methods). The model generates networks with different structures by controlling the fitness parameter $\beta$ (see Methods). The average degree $\langle \kappa \rangle$ is fixed for all networks, but the average betweenness $\langle b \rangle$ and average clustering coefficient $\langle cc \rangle$ (i.e.\ triangles) are larger than expected in the configuration model (fixed degree sequence and rewired social ties~\cite{fosdick2018configuring}) (Table~\ref{tab:01}). Figure~\ref{fig04} shows that lower $\beta$ generates networks closer to those produced by the standard preferential attachment model (BA model), whereas larger $\beta$ results in networks with fewer hubs, higher betweenness, and higher clustering coefficient. A modular structure emerges (measured by the modularity $\langle Q \rangle$~\cite{newman2006modularity}) for larger $\beta$, but the networks are not assortative by degree (measured by the assortativity index $r$~\cite{newman2018networks}), independently of the value of $\beta$. These results show that the fitness function (based on attribute preference) creates clustering and, consequently, brokerage in the network.

\begin{table}[t]
\centering
\begin{tabular}{ccccccc}
\hline
model & $\beta$ & $\langle k \rangle$ & $\langle b \rangle$ & $\langle cc \rangle$ & $\langle Q \rangle$ & $\langle r \rangle$ \\ \hline
fitness       & 0   & 5.98  & 0.0025 & 0.029 & 0    & -0.08 \\
configuration & -   & 5.98  & 0.0025 & 0.025 & 0    & -0.21 \\\hline
fitness       & 20  & 5.98  & 0.0030 & 0.086 & 0.57 & -0.09 \\
configuration & -   & 5.98  & 0.0026 & 0.021 & 0    & -0.24 \\\hline
fitness       & 100 & 5.98  & 0.0036 & 0.176 & 0.69 & -0.06 \\
configuration & -   & 5.98  & 0.0027 & 0.014 & 0    & -0.29 \\\hline
Watts-Strogatz& -   & 6     & 0.0036 & 0.177 & 0.50 & -0.04 \\
configuration & -   & 6     & 0.0032 & 0.003 & 0    & -0.01 \\\hline
\end{tabular}
\caption{Summary of network statistics for the fitness random network model and the respective configuration model (swapping of one of the ends of two social ties chosen at random~\cite{fosdick2018configuring}). } 
\label{tab:01}
\end{table}

\begin{figure*}[!th]
    \centering
    \includegraphics[scale=0.6]{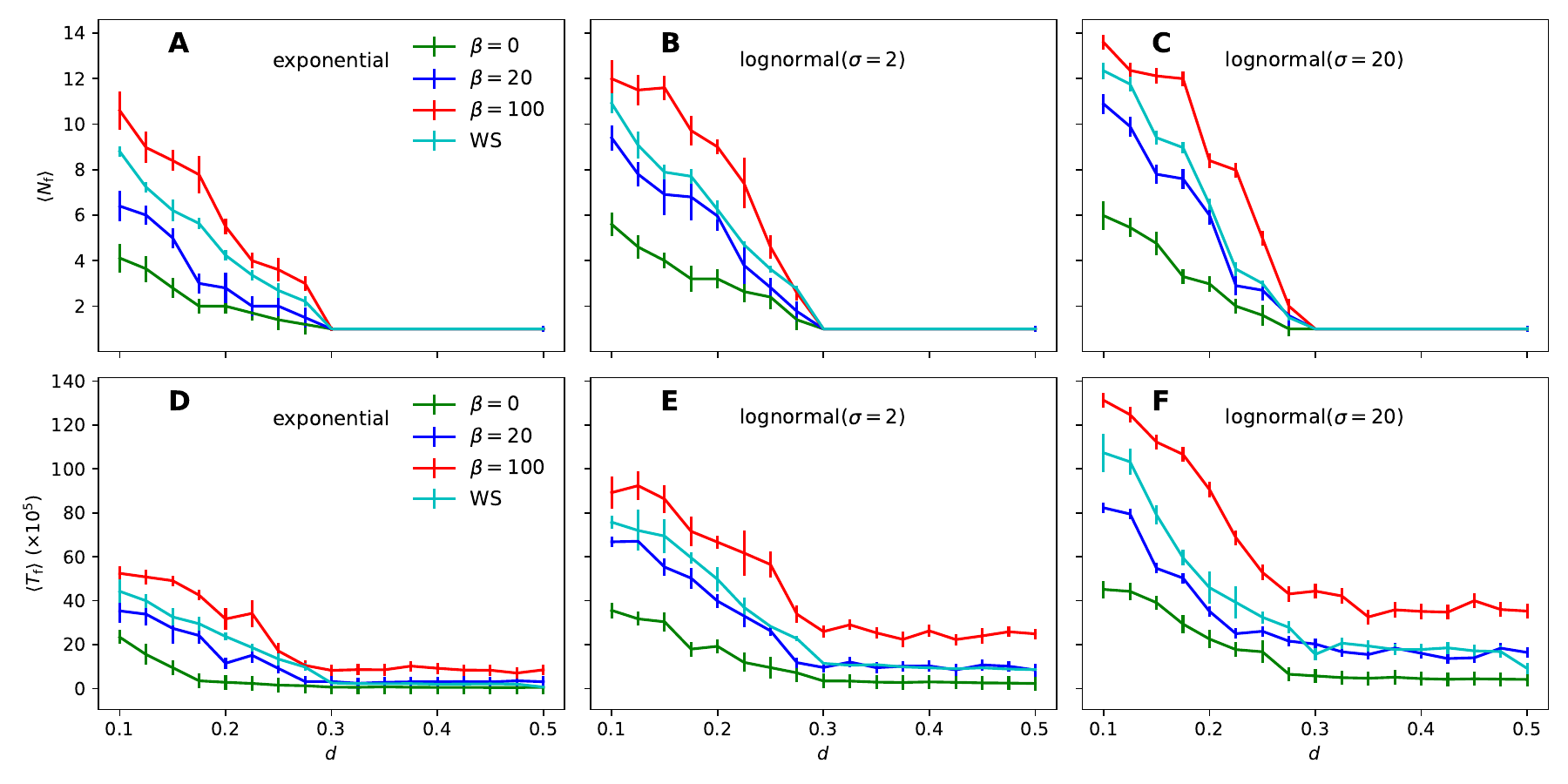}
    \centering\caption{The average number of opinion clusters ($\langle N_\text{f} \rangle$) for the (A) exponential, (B) log-normal ($\sigma=2.7$), and (C) log-normal ($\sigma=20$) activation times as a function of the confidence level $d$. The average stabilisation time ($\langle T_\text{f} \rangle$) for the (D) exponential, (E) log-normal ($\sigma=2.7$), and (F) log-normal ($\sigma=20$) activation times as a function of the confidence level $d$. The averages are taken over $m=50$ realisations. Vertical bars represent standard errors. }
    \label{fig05}
\end{figure*}

Figure~\ref{fig05}A shows that the clustered structure (increasing $\beta$) affects the opinion dynamics by creating more opinion clusters for $d<0.3$, whereas it has no effect for $d \geq 0.3$ in the stationary state. Adding the heterogeneous temporal activity further affects the opinion dynamics by increasing the number of opinion clusters for all values of $\beta$ (Fig.~\ref{fig05}B,C). The results also indicate an overall slowdown of the opinion dynamics towards stationarity (Fig.~\ref{fig04}E-F). In this case, the slowdown occurs for all values of $d$ but is more pronounced for $d<0.3$ (when opinion clusters emerge). The effect of burstiness is relatively more significant on the convergence time ($T_{\text{f}}$) compared to the formation (number) of opinion clusters ($N_{\text{f}}$). The triangles are sufficient to increase diversity (WS model), but the effect increases in the fitness model with the same average clustering coefficient but higher modularity (See Table~ \ref{tab:01}).

\begin{figure*}[!th]
    \centering
    \includegraphics[scale=0.5]{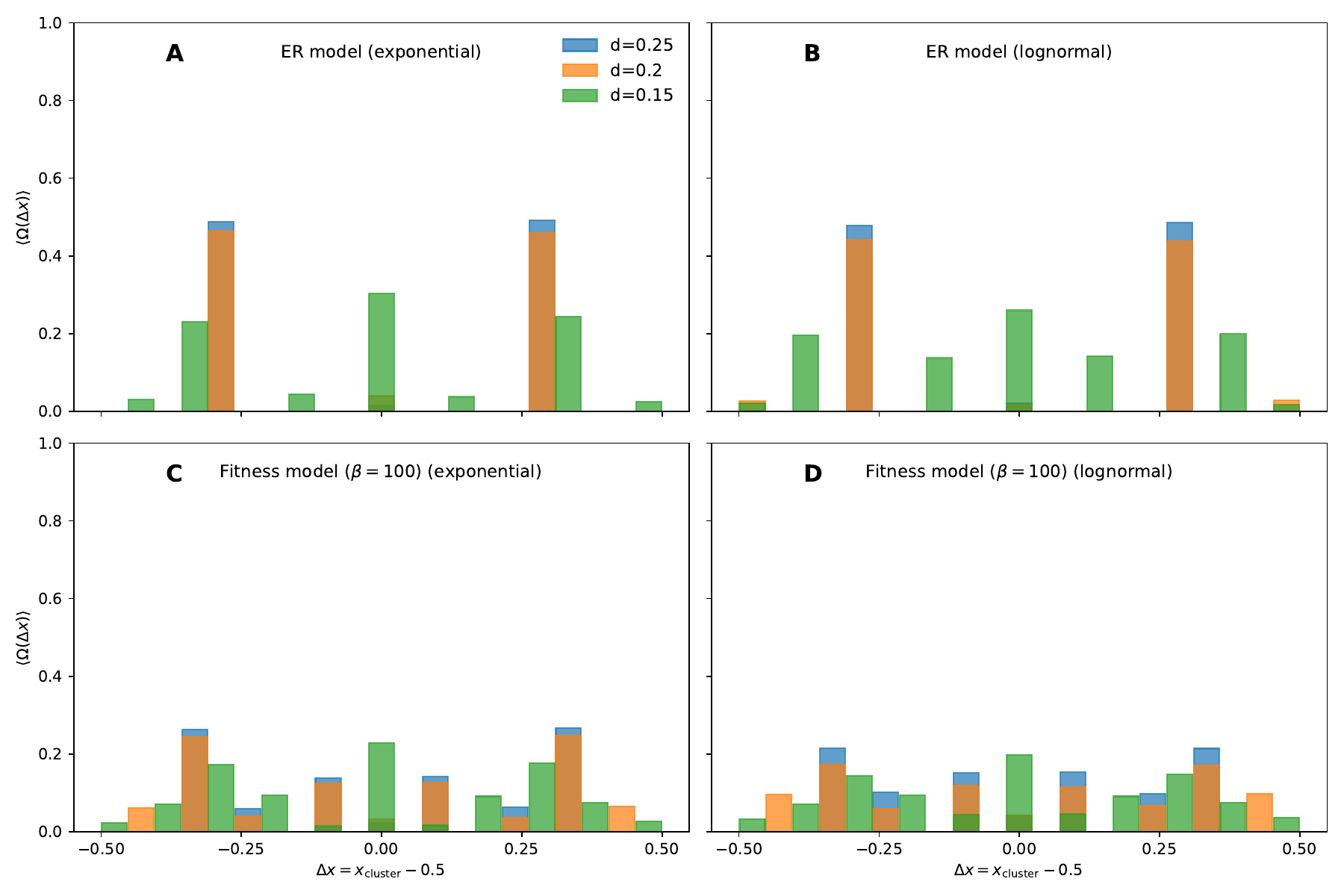}
    \centering\caption{The relative average size of an opinion cluster ($\langle \Omega (\Delta x) \rangle$) as a function of the distance ($\Delta x = x_{\text{cluster}}-0.5$) of the cluster from the central opinion ($x=0.5$) for various confidence levels $d$. (A) ER model and exponential activation times, (B) ER model and log-normal ($\sigma=2.7$) times, (C) fitness model and exponential times, and (D) fitness model and log-normal times ($\beta=100$). Simulations are repeated for $m=50$ starting conditions with the same set of parameters.}
    \label{fig06}
\end{figure*}

The results show that network clustering increases the emergence of opinion clusters while slowing down the creation process of those clusters. Network clustering, particularly triangles, creates local redundancies in communication and thus locally reinforces existing opinions. On the other hand, bridging nodes are exposed to opinions from different clusters and therefore take longer to converge to one or another opinion. The introduction of burstiness on clustered networks boosts the reinforcement mechanism because bursts on redundant social ties (e.g.\ in triangles) promote exchanges between social neighbours. This leads to local consensus and makes the small opinion clusters too different early in the dynamics. After a few interactions, individuals struggle to find sufficiently close neighbours (in the opinion space) to potentially adjust opinions. Furthermore, the long inter-event times slow the convergence to the stationary state because they counter-balance the fast local convergence with individuals having intermediate opinions.

The confidence level $d$ and structural and temporal patterns affect the number of opinion clusters, as seen in Figure~\ref{fig05}A-C. We now measure the relative size of these opinion clusters ($\Omega$), i.e.\ the number of individuals per cluster divided by $N$. Figure~\ref{fig06} shows that opinion clusters are symmetric around the centre of the opinion space and $\langle \Omega \rangle$ decreases with the increasing number of opinion clusters $N_{\text{f}}$. This means that the individuals are redistributed to different opinion clusters when more clusters emerge in the opinion dynamics. In configurations where individuals self-organise into several opinion clusters (e.g.\ for smaller $d$), cluster size heterogeneity exists. Still, the emergence of (large or small) disproportionate clusters is not observed. Opinion clusters are usually equally spaced but not always of the same size in the (structure and activity) homogeneous case (Fig.~\ref{fig06}A). Adding temporal heterogeneity brings a better balance on the size of clusters in both homogeneous (ER)(Fig.~\ref{fig06}B) and heterogeneous (fitness)(Fig.~\ref{fig06}D) network structures. Furthermore, extreme opinions are more likely when multiple clusters emerge, i.e.\ in case of higher diversity, but they are generally observed within a relatively small fraction of the population.

\section{Discussion and Conclusions}

Individuals attempt to adjust their opinions via social interactions and communication. One-to-one exchanges occur when individuals are similar, but those with disparate opinions hardly reach consensus~\cite{baronchelli2018emergence}. The cost to reduce large gaps in the opinions of different individuals is too large, and potential incremental gains may be relatively insignificant and unable to homogenise opinions. Attempting to mitigate large gaps may be more damaging because lacking common ground may encourage individuals to push back if facing opposite views~\cite{bail2018exposure}.

In this paper, we approached the problem from a different perspective. We studied basic mechanisms to encourage interactions between relatively similar individuals and analysed the potential for departing from consensus and polarisation towards a more diversified opinion landscape. We designed a temporal version of the Deffuant opinion dynamics model to include the social network structure and temporal patterns of pairwise interactions. We first studied the impact of burstiness on the formation of opinion clusters (local consensus), then analysed the effect of the same temporal patterns in combination with social network clustering on opinion dynamics.

A fundamental mechanism to advance opinions towards others is reinforcement, the act of repeatedly exposing someone to a given opinion so that the individual will eventually accept it. Reinforcement may be implemented by social influence of peers (i.e.\ social contacts)~\cite{guilbeault2018complex}, nudging~\cite{Thaler2008}, or, as we demonstrated in this paper, by bursts of interactions. We found that burstiness promotes local consensus due to frequent pairwise opinion adjustments during short periods, followed by longer periods of no interaction, slowing down the process. The emergence of various spots of local consensus impedes individuals to self-organise into larger groups sharing the same opinion. Homogeneous (Poisson) temporal activity often leads to consensus or the emergence of two or four opinion clusters, typical signs of polarisation. burstiness, on the other hand, leads to a multi-partisan population with up to ten opinion clusters in completely random social structures. As expected, this effect is more substantial for lower tolerance to adjust opinions; if individuals only accept interacting with those very similar, chances are higher that multiple opinion clusters emerge. As mentioned above, local structural clustering (i.e.\ social triangles) also promotes reinforcement. The higher the clustering, the higher the number of opinion clusters. According to our computational experiments, the combination of burstiness and clustering substantially reduces polarisation with the emergence of at least ten opinion clusters for relatively high tolerance values ($d \leq 0.2$). Furthermore, there is no emergence of disproportionately large clusters dominating the population, independently of the number of opinion clusters. On the other hand, burstiness promotes a more balanced cluster configuration. Moderate opinions generally prevail, but diversity also leads to higher chances of extreme opinions, albeit those clusters with extreme opinions are relatively small.

These results suggest that homogeneity (i.e.\ a lack of social clustering and absence of burstiness) promotes polarisation. burstiness and structural clustering (triangles and community structure), on the other hand, increase opinion diversity. A maximum variety of opinions is, of course, observed if each individual has their own opinion. However, in real settings, groups often converge to a consensus by sharing similar views. Fitting our model to real data is challenging because data on the temporal patterns of daily offline social interactions are unavailable. However, previous research has shown that the distribution of inter-event times of communication via letters fits a power-law $P(\Delta t) \propto \Delta t^{-1.5}$~\cite{oliveira2005} (which is approximately our log-normal model with $\sigma=2.7$) whereas online communication follows $P(\Delta t) \propto \Delta t^{-1}$~\cite{barabasi2005origin} (which is approximately our log-normal model with $\sigma=20$). In these cases, our experiments suggest that online communication should create less polarisation than expected in offline communication. We argue that this phenomenon is not observed in real-life (i.e.\ less polarization online) because structural clustering is substantially lower in online social media ($cc_{\text{median}} \sim 0.04$) compared to offline social networks ($cc_{\text{median}} \sim 0.37$)~\cite{rocha2021}. We stress that the relatively low social clustering observed in social media is sufficient to promote polarisation. The reported difference in social clustering is independent of the network size~\cite{rocha2021}, but the studied offline social networks are relatively small, with a few hundred individuals. Individuals are constrained and organised into strong (network) communities (e.g., within schools, work, clubs, cities). The potential fragmentation of offline social networks would further promote diversification and a multi-partisan society, as suggested by our model. This rationale implies that the low cost of creating online friends, leading to almost unrestricted opportunities for social interactions, creates sufficient conditions for polarisation, independently of any other mechanisms.

Our study also showed that the convergence towards stationarity is longer in the presence of heterogeneity, with both network clustering and burstiness contributing to the slowdown of the dynamics. This is a consequence of the coexistence of several small clusters of different opinions. Bridging individuals (i.e.\ brokers, with high betweenness and more often observed in highly clustered networks) require more time to position themselves in a particular opinion cluster because they are exposed to different social groups. Similarly, the longer periods of absence of activity (due to the tail of the distribution of inter-event times) increases the time needed for pairs of individuals to converge to one or another opinion~\cite{delvenne2015diffusion, jo2021, li2023}. Similar dynamics have been observed in the spread of simulated infectious diseases, with bursts initially accelerating the spread but eventually slowing down the diffusion because the long periods of inactivity reduce the chances of finding susceptible nodes to infect~\cite{rocha2011, rocha2013bursts, karsai2011, stehle2011simulation}.

Our study starts with the assumption that individuals adjust their opinions via pairwise interactions with similar peers. Therefore, both social network structure and the timings of social interaction are shown to be responsible for regulating the emergence of groups of individuals sharing the same opinion. This is reasonable in both offline and online environments. However, in real-life settings, individuals are also exposed to mass media (e.g.,\ radio, TV, newspapers) and group activism that may contribute to shaping opinions. Future models could include broadcasting and synchronous group interactions to accommodate that mechanism~\cite{Hegselmann2002}. Information personalisation also plays a role. In offline social interactions, power relations, norms, or contextual conversations may prevent individuals from sharing certain opinions socially. On the other hand, automated algorithmic personalisation is the norm in online settings and biases the information exposed to individuals based on their previous activity, creating the so-called filter bubbles~\cite{perra2019modelling}. Although all those mechanisms contribute to opinion dynamics and can be actively exploited to manipulate opinions, our study suggests that more attention must be given to reshaping the structure of online social networks to promote a multi-partisan society. Counter-intuitively, burstiness in social media communication may induce diversity by creating local spots of consensus. Such diversity may be boosted by increasing social redundancy via social triangles to strengthen online social relations.

\bibliographystyle{ieeetr}
\bibliography{References}

\end{document}